\documentclass{WileyMSP-template}
\usepackage{colortbl}   
\usepackage{xcolor}     
\usepackage{steinmetz}
\graphicspath{{./images/}}
\usepackage{gensymb}
\usepackage{ragged2e} 
\usepackage{amsmath}
\usepackage{booktabs}
\begin{document}
\justifying

\pagestyle{fancy}

\title{Efficient magnetization switching driven by orbital torque originating from light 3d-transition-metal nitrides}

\maketitle


\author{Gaurav K. Shukla$^{1,*}$}
\author{Yoshio Miura$^{2}$}
\author{Mayank K. Singh$^{1}$}
\author{Shinji Isogami$^{1}$}


$^{1}$Research Center for Magnetic and Spintronic Materials (CMSM),  
National Institute for Materials Science (NIMS), Tsukuba 305-0047, Japan

$^{2}$Faculty of Electrical Engineering and Electronics,  
Kyoto Institute of Technology, Kyoto 606-8585, Japan

Email:shukla.gauravkumar@nims.go.jp



\begin{abstract}
The orbital Hall effect (OHE) in light transition metals offers a promising route to generate orbital torques for efficient magnetization control, providing an alternative to conventional spin Hall effect approaches that rely on heavy metals. We demonstrate perpendicular magnetization switching in [Co/Pt]$_3$ multilayers driven by the OHE in a light 3d transition metal nitride, VN, with 111-texture of face-center cubic structure. Second harmonic Hall measurement reveals a large torque efficiency of $\sim$ -0.41 in the VN(7.5 nm)/[Co(0.35 nm)/Pt(0.3 nm)]$_3$, which significantly surpasses that in the control samples with Co, Py, and CoFeB ferromagnets, suggesting strong conversion of orbital current originating from VN to spin current by [Co/Pt]$_3$ ferromagnet. Full switching by in-plane current is achieved with an in-plane magnetic field, while partial field-free switching occurs without it. The critical current density for the switching is found to be comparable to that of the W-based spin-orbit torque device. First-principles calculations confirm a large orbital Hall conductivity in VN, with a small spin Hall conductivity around the Fermi energy. Our results highlight the potential in the combination of light 3d transition metal nitrides and Co/Pt ferromagnetic multilayer with 111-texture to maximize the magnetization switching efficiency of orbitronic devices.

\end{abstract}

\section*{Introduction}
Spin–orbit torque (SOT)-based devices enable electrical control of magnetization switching through the spin Hall effect (SHE) \cite{shao2021roadmap}. Most studies have focused on heavy metals such as Pt, Ta, and W, where strong spin–orbit coupling (SOC) produces a large spin Hall angle ($\theta_{SH}$) and efficient spin-current generation \cite{sato2020cmos,bekele2021high}. Among them, $\beta$-W is regarded as one of the promising candidates for large-scale SOT-MRAM integration owing to its $\theta_{SH}$ $\sim$ 0.3-0.4 \cite{pai2012spin}, while its high resistivity leads to large power consumption\cite{gupta2025harnessing}. Platinum (Pt), on the other hand, offers a favorable compromise of moderate resistivity ($\sim$50 \textmu$\Omega$\,cm) and $\theta_{SH}$$\sim$0.08, making it the most widely used spin source \cite{aradhya2016nanosecond,nguyen2016spin}.

\begin{figure}[t]
   \centering
   \includegraphics[width=0.5\textwidth]{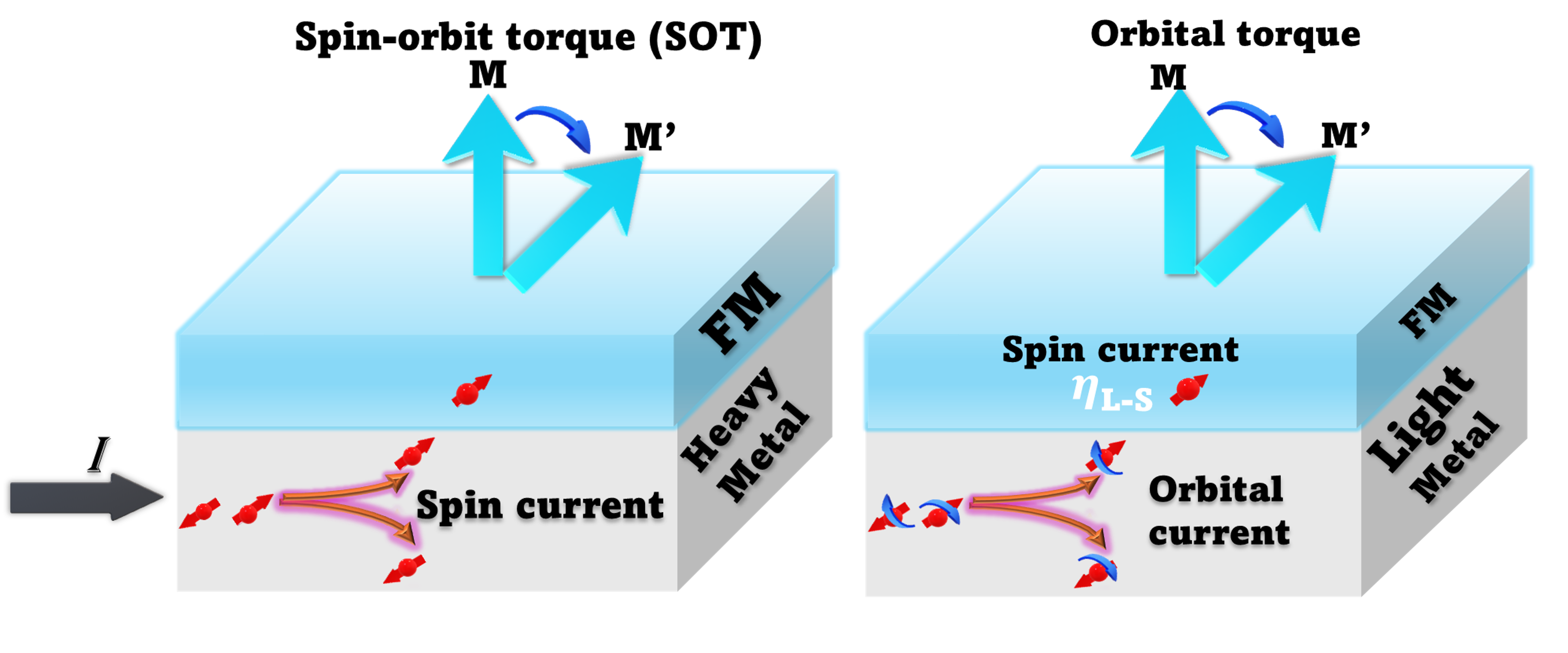}
   \caption{Schematic diagram of (a) spin-orbit torque (SOT) through spin Hall effect (SHE) and (b) Orbital torque via orbital Hall effect (OHE).}
    \label{Fig1}
\end{figure}

Recently, current-driven magnetization switching has also been observed in devices without heavy metals, i.e., the light-metal (LM)/ferromagnet (FM) bilayer structures, even though the SOC of LMs is much weaker than that of heavy metals \cite{ando2025orbitronics,lee2021efficient,yang2024orbital}. The origin of torque observed in the FM layer is different from the conventional SHE, which is referred to as the orbital-Hall effect (OHE) in the LM that originates from the orbital textures in momentum space \cite{go2020orbital,choi2023observation,fukunaga2023orbital,zhao2025efficient}. This discovery has opened the field of modern orbitronics, where magnetization of FM can be controlled by orbital currents from LM layers, offering a new materials-design pathway for SOT devices beyond conventional heavy metals \cite{choi2023observation,fukunaga2023orbital,zhao2025efficient}. The torque efficiency in the LM/FM bilayers is characterized by the effective spin-Hall conductivity (SHC, $\sigma^{eff}_{SH}$) given as \cite{go2020orbital}
\begingroup
\setlength{\abovedisplayskip}{5pt}
\setlength{\belowdisplayskip}{5pt}
\begin{equation}\label{eq:A1}
    \sigma^{eff}_{SH} = \sigma^{LM}_{SH} + \eta_{L-S}\sigma^{LM}_{OH}
\end{equation}
\endgroup
where $\sigma_{SH(OH)}$ and $\eta_{L-S}$ represent SHC (orbital Hall conductivity (OHC)) and conversion efficiency from orbital angular momentum ($L$) to spin angular momentum ($S$), respectively. An enhanced $\sigma^{eff}_{SH}$ is desired to achieve efficient SOT devices; therefore, fundamental studies have been extensively conducted to achieve a large $\sigma^{LM}_{SH}$ (first term of Eq.\ref{eq:A1}) and $\eta_{L-S}\sigma^{LM}_{OH}$ (second term of Eq.\ref{eq:A1}). The LMs such as Ti, V, and Cr are focused due to extremely high $\sigma^{LM}_{OH}$ to date, which is 10$^1$$\sim$10$^3$ times higher than the $\sigma^{LM}_{SH}$ \cite{lee2021orbital,sala2022giant,wang2025orbitronics,liu2024harnessing}, promising the enhanced $\sigma^{eff}_{SH}$ due to the large second term of Eq.\ref{eq:A1} compared with the first term \cite{choi2023observation,zhao2025efficient}. However, the orbital currents do not directly act on the magnetization of the FM layer; they offer significant potential to generate large spin currents via efficient $L-S$ conversion, as shown in Figs.\ref{Fig1}(a) and Figs.\ref{Fig1}(b). This $L-S$ conversion arises due to the SOC of FM material \cite{zhao2025efficient}. For instance, in one of the LM/FM bilayer systems, Cr/Co, replacing the Co layer with Gd results in a tenfold increase in torque efficiency \cite{lee2021efficient}. This enhancement is attributed to the higher $\eta_{L-S}$, despite the equally small SOC of LM. Therefore, material development of FM with large $\eta_{L-S}$ is indispensable, in addition to the LM with large $\sigma^{LM}_{OH}$. Furthermore, a perpendicularly magnetized FM layer and its field-free switching are also desired for SOT-devices in terms of scalability and high-speed memory functionality \cite{fukami2016spin}. Since fundamental works on the $\eta_{L-S}$, $\sigma^{LM}_{OH}$, and field-free switching have been conducted individually to date, enhancing the efficiency, which is the next major challenge for orbitronics. In this study, therefore, we aim to develop another LM with sufficient $\sigma^{LM}_{OH}$ and optimum interface of LM/FM bilayer structure to maximize the SOT efficiency, maintaining the field-free perpendicular magnetization switching.

In the quest for efficient orbitronic platforms, nitrogen (N)-based materials have recently attracted attention. These systems exhibit enhanced damping-like torque efficiency \cite{tripathi2024impact}, field-free magnetization switching \cite{kumar2025unconventional}, large OHE \cite{kumar2025unconventional}, Berry-curvature-driven magnetization switching \cite{shukla2025berry}, and long spin diffusion lengths \cite{swatek2022room}. Their metallic character, originating from d–p orbital hybridization, together with excellent chemical stability, ensures compatibility with device integration \cite{bi2025rise,isogami2023antiperovskite}. These attributes position transition-metal nitrides as promising candidates for next-generation orbitronic devices capable of deterministic and energy-efficient magnetization control. 
 
 Here, we employed (111) plane-oriented VN with face-centered cubic (FCC) structure as the LM layer to explore the advantage for significant $\sigma^{LM}_{OH}$, comparing with its spin-counterpart. As for the FM layer with perpendicular magnetization, we employed [Co/Pt]$_3$ multilayer structure, for two reasons, which are the favorable crystal growth on the (111)-VN layer and the presence of Pt with sufficient SOC. Therefore, seamless integration of [Co/Pt]$_3$ on (111)-VN enables robust perpendicular magnetization together with strong $\eta_{L-S}$. Using such a fully (111)-oriented VN/[Co/Pt]$_3$ bilayer structure, the second-harmonic Hall measurements were carried out. As a result, an orbital torque efficiency was estimated to be –0.16 $\sim$ –0.41 as the VN thickness increases, while much smaller efficiencies (0.08, 0.015, and 0.002) were estimated in the control samples replacing the [Co/Pt]$_3$ with Co, Py, and CoFeB, respectively. This reduction in torque effciency suggest the strong conversion of orbital current originating from VN to spin current by [Co/Pt]$_3$ FM layer. The full magnetization switching in VN/[Co/Pt]$_3$ is observed in the presence of an in-plane external magnetic field, while partial field-free switching is achieved without it. The partial field-free switching is attributed to the tilted magnetization of [Co/Pt]$_3$ FM layer that breaks mirror symmetry normal to the film plane. The power density of the VN/[Co/Pt]$_3$ is found to be four times lower than that of controlled W/[Co/Pt]$_3$, suggesting energy-efficient switching by VN. Our findings highlight the promising bilayer structures for orbital-based SOT-devices, owing to the hybrid effects of fully (111)-oriented transition-metal nitrides and [Co/Pt]$_3$ ferromagnetic multilayers.
\section*{Results}
\subsection*{Structural analysis}
\begin{figure}[htbp]
   \centering
   \includegraphics[width=0.5\textwidth]{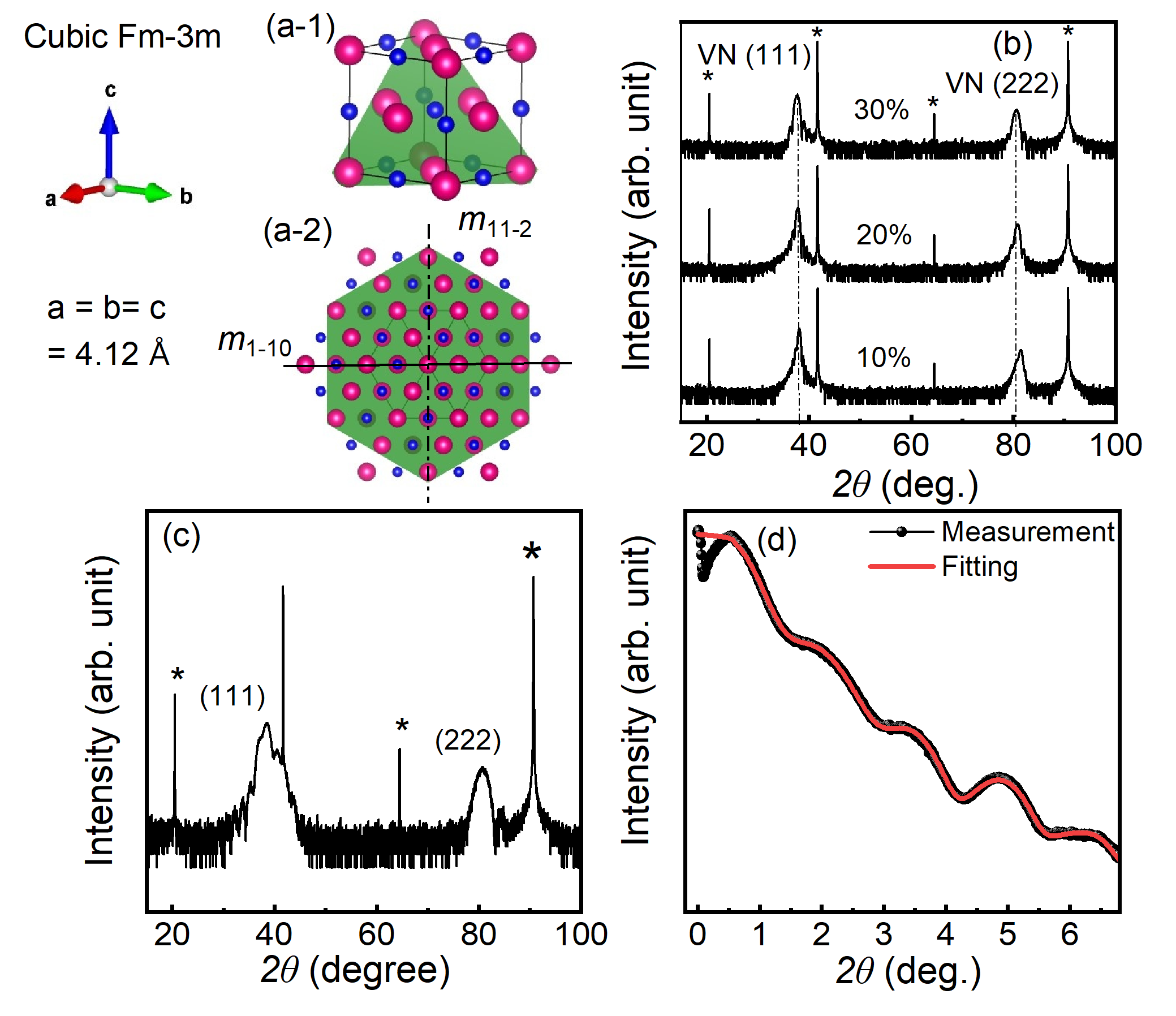}
   \caption{ (a-1) A face-centered (FCC) unit cell of VN, where pink and blue color spheres represent the V and N atoms, respectively. (a-2) The (111) plane view of the crystal. The solid line represents the preserved mirror symmetry along [1-10] direction, while the dashed line represents the broken mirror symmetry along [11-2] direction. (b) X-ray diffraction (XRD) profile for the Al$_2$O$_3$ substrate//VN(5\,nm) with different N$_2$ gas flow ratio relative to Ar gas during sputter deposition process. The (*) denotes the substrate peaks. (c) XRD profile of the Al$_2$O$_3$ substrate//VN(5\,nm)/[Co(0.35\,nm)Pt(0.3\,nm)]$_3$/MgO(3\,nm) multilayer stack. (d) The X-ray reflectivity (XRR) data (black color) and fitted curve (red color) for the same multilayer stack.}
    \label{Fig2}
\end{figure}

Figure\,\ref{Fig2}(a-1) shows the unit cell of VN with space group Fm$\bar3$m, in which the Wyckoff's positions (0,0,0) and (0.5,0.5,0.5) are for the V and N atoms, respectively. Figure\,\ref{Fig2}(a-2) shows the (111)-plane of the VN supercell, where mirror symmetry is retained along [1-10] but broken along [11-2] direction. Figure\,\ref{Fig2}(b) shows the X-ray diffraction (XRD) profiles for the 5nm-thick VN films grown on C-plane oriented Al$_2$O$_3$ substrates with various N$_2$ flow ratios relative to the Ar gas during the reactive sputtering process. XRD peaks were observed at $2\theta$ $\sim$ 38$\degree$ and 80.6$\degree$ originating from the (111)- and (222)-plane of VN, respectively, for all N$_2$ ratios. The results suggest the robust (111)-oriented epitaxial growth of VN films on the Al$_2$O$_3$ substrates. The out-of-plane lattice constant ($c$) was estimated to be 4.12\,\AA. 
 \begin{figure*}[t]
   \centering
   \includegraphics[width=0.9\textwidth]{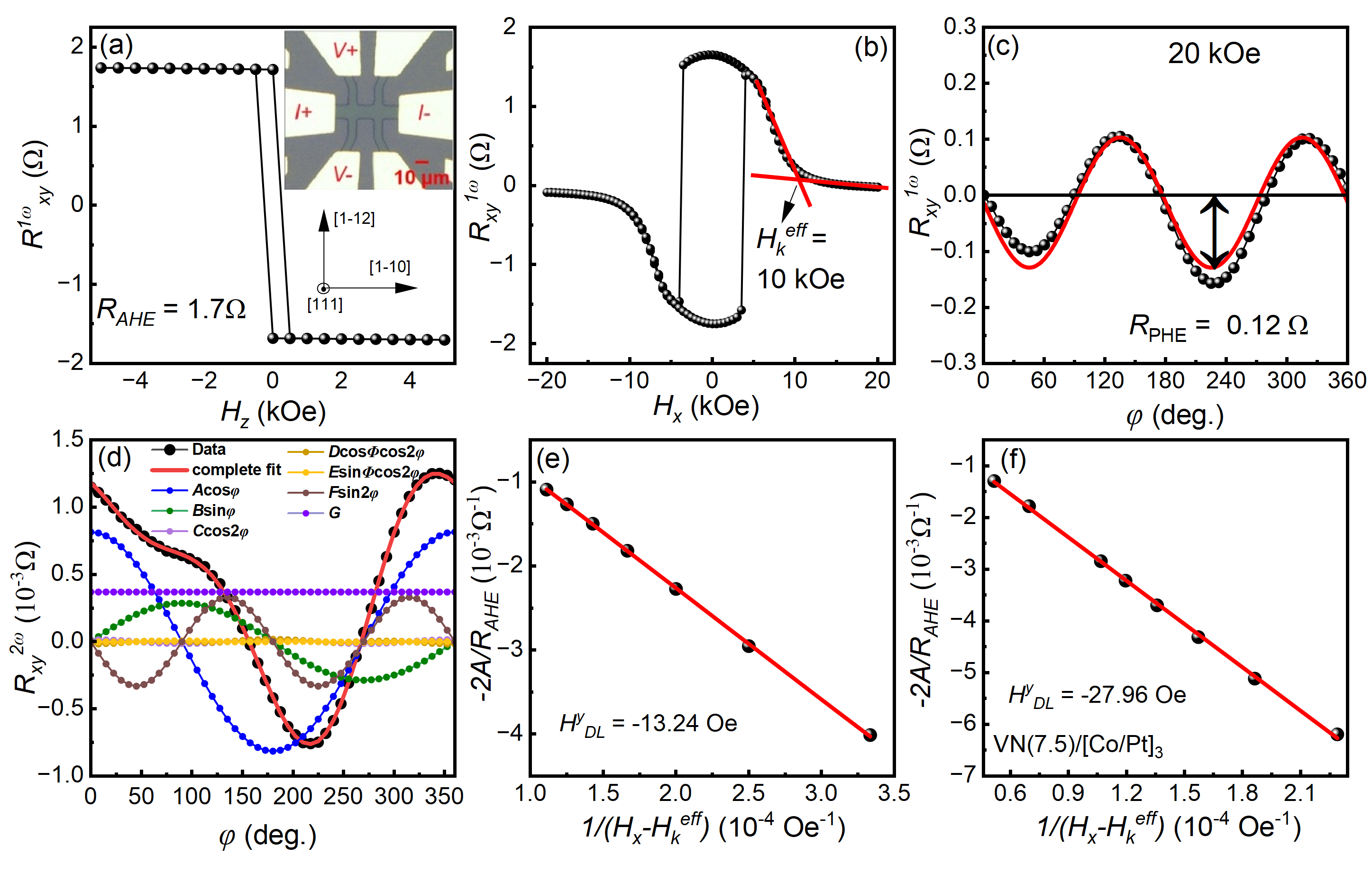}
   \caption{\justifying (a,b) Dependence of first harmonic Hall resistance \(R_{xy}^{1\omega}\) for the Al$_2$O$_3$ substrate//VN(5 nm)/[Co(0.35 nm)/Pt(0.3 nm)]$_3$/MgO(3 nm) on the magnetic field along out-of-plane direction (\(H_z\)) (a) and in-plane direction that is parallel to the current channel ($H_x$) (b). The inset photograph shows a representative microfabricated Hall device. (c) Dependence of the \(R_{xy}^{1\omega}\) on the azimuthal angle ($\varphi$) of $H_x$ with respect to current channel. (d) same as (c) but for second harmonic Hall resistance \(R_{xy}^{2\omega}\). The fits Eq.\,\ref{eq:2} are shown in different colors, while the red curve represents the total of each component. (e) Plot of the component \(-2A/ R_{\mathrm{AHE}}\) versus \(\frac{1}{H_x - H_k^{\mathrm{eff}}}\) and linear fitting is in red. (f)  Same as Fig.\ref{Fig3}(e) but for VN(7.5\,nm)[Co/Pt]$_3$.}
\label{Fig3}
\end{figure*}
\begin{figure*}[t]
   \centering
   \includegraphics[width=0.9\textwidth]{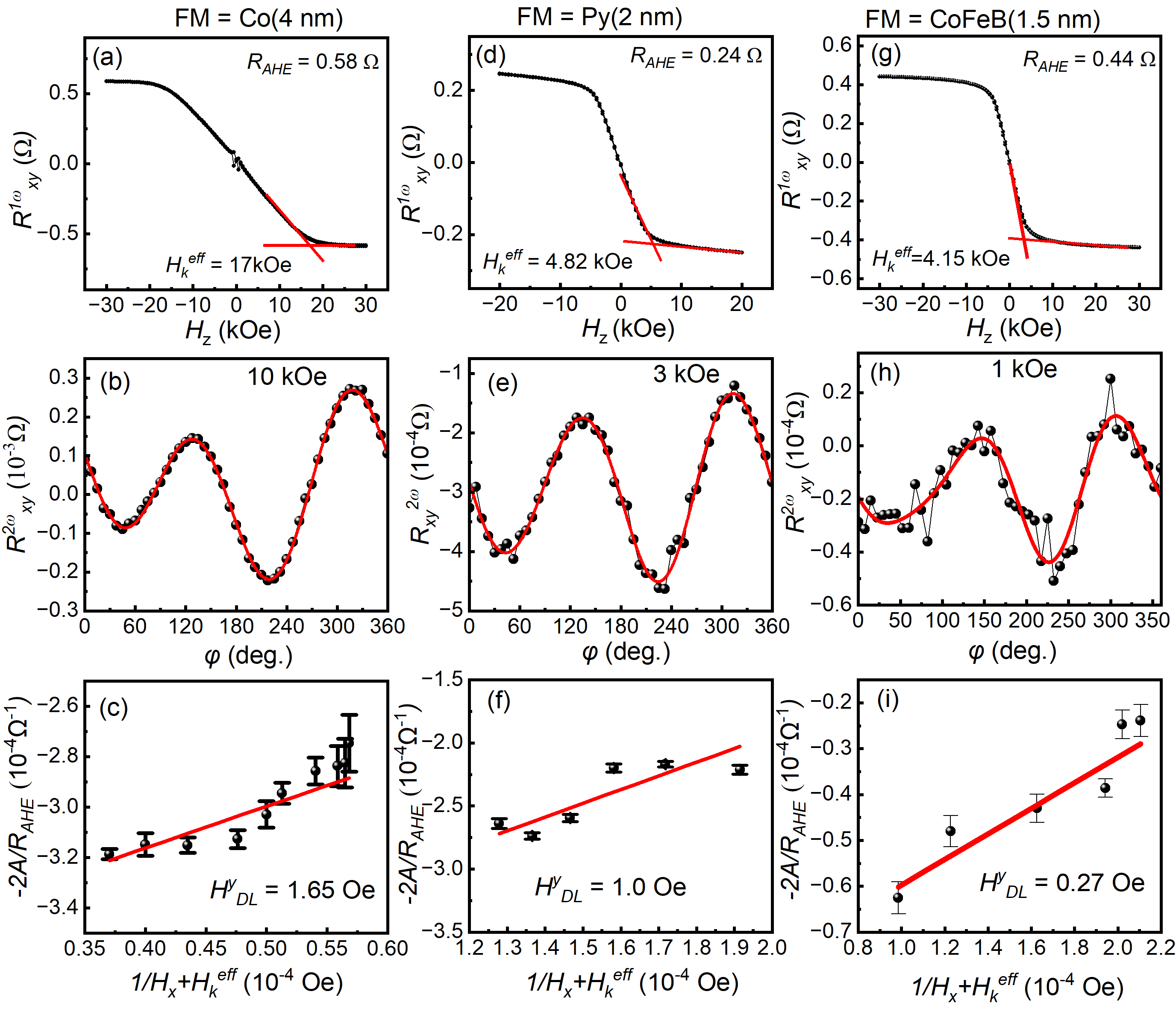}
   \caption{(a, b, c) Out-of-plane magnetic field ($H_z$) dependent of $R_{xy}^{1\omega}$ (a), in-plane azimuthal angle ($\varphi$) dependence of $R_{xy}^{2\omega}$ with respect to the current channel and fitting using Eq.\ref{eq:2} is in red (b),  -$2A/ R_{\mathrm{AHE}}$ versus $\frac{1}{H_x + H_k^{\mathrm{eff}}}$ with linear fitting (c) for the VN (5\,nm)/Co(4\,nm). (d,e,f) Same as Figs.\ref{Fig4} (a,b,c) but for VN(5\,nm)/Py(2\,nm). (g,h,i) Same as Figs.\ref{Fig4} (a,b,c) but for VN(5 nm)/CoFeB(1.5\,nm).}
    \label{Fig4}
\end{figure*}
 \begin{figure}[t]
   \centering
  \includegraphics[width=0.5\textwidth]{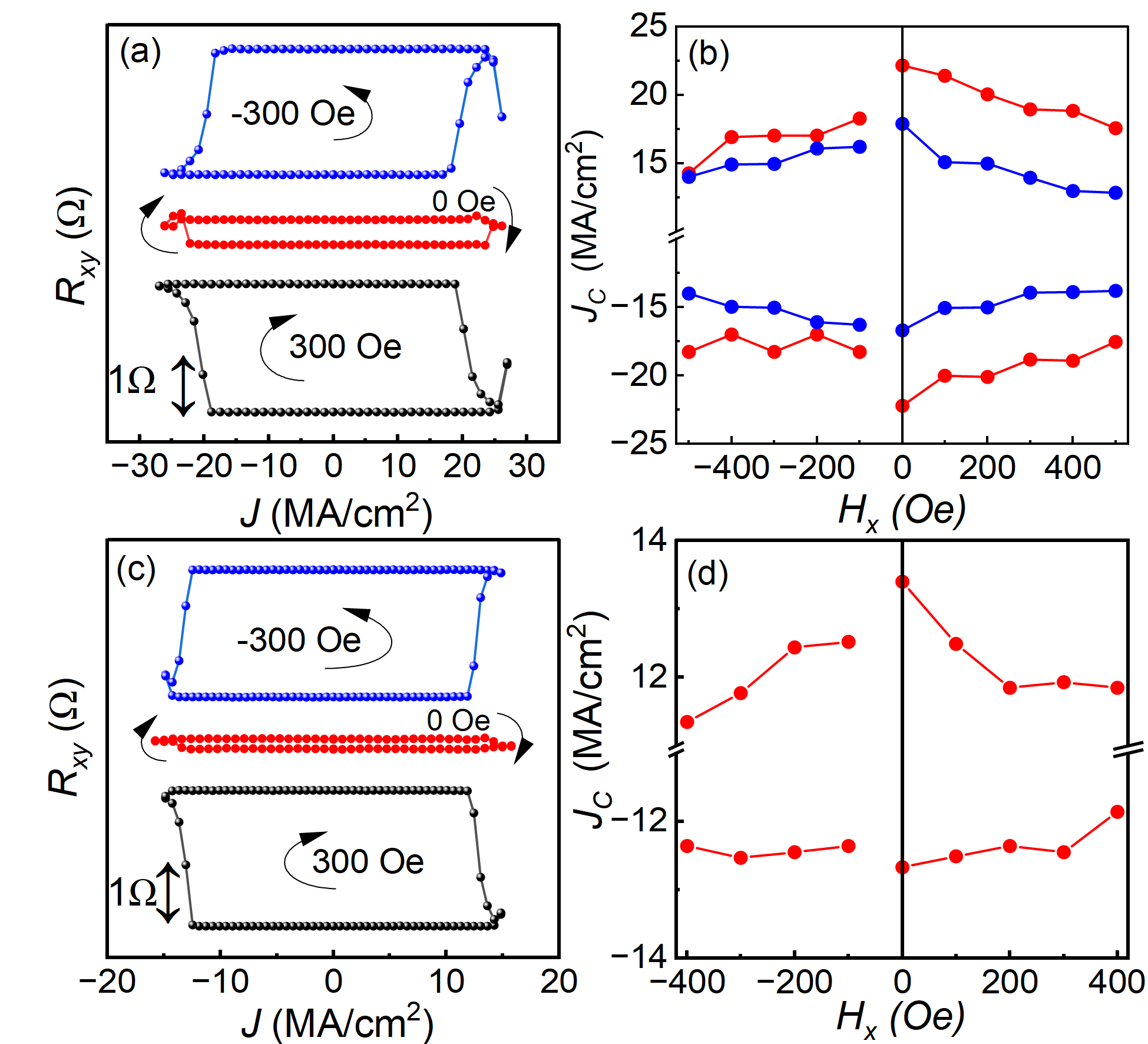}
  \caption{(a) Current-induced magnetization switching (CIMS) hsyerisis loop with various in-plane magnetic fields ($H_x$) for the main sample Al$_2$O$_3$ substrate//VN(5 nm)/[Co(0.35 nm)Pt(0.3 nm)]$_3$/MgO(3 nm). (b) Relationship between the critical current density ($J_c$) and $H_x$ for the main sample (red) and the Al$_2$O$_3$ substrate//VN(7.5 nm)/[Co(0.35 nm)Pt(0.3 nm)]$_3$/MgO(3 nm) (blue). (c,d) Same data as Fig.5\,(a,b), but for control sample Al$_2$O$_3$ substrate//W(5 nm)/[Co(0.35 nm)/Pt(0.3 nm)]$_3$/MgO(3\,nm).}
    \label{Fig5}
\end{figure}     
Figure\,\ref{Fig2}(c) shows the XRD profile for the main sample in this study: Al$_2$O$_3$ substrate//VN(5 nm)/[Co(0.35 nm)/Pt(0.3 nm)]$_3$/MgO(3 nm) (hereafter referred to as VN(5)/[Co/Pt]$_3$). In addition to the same XRD peaks originating from the VN layer with (111) orientation (Fig.\,\ref{Fig2}(b)), another XRD peak appeared near the peak of (111)-VN can be attributed to the [Co/Pt]$_3$ layer on the VN. The possible crystal structures of the [Co/Pt]$_3$ layer are (111)-oriented face-centered cubic (FCC) or C-plane oriented hexagonal, and their mixture, the separation for which is difficult for the out-of-plane XRD because of equivalent crystal symmetry between the (111)-plane and the C-plane. However, note that we confirm the epitaxial growth of [Co/Pt]$_3$ on the (111)-VN layer. Figure\,\ref{Fig2}(d) presents the X-ray reflectivity (XRR) profile (black) and corresponding fit (red) for the same stacking as Fig.\,\ref{Fig2}(c). It was revealed that the interlayer mixing and/or alloying between the VN and [Co/Pt]$_3$ layers can be ignored (see Supplementary Note 1). 
 \subsection*{Spin-orbital torque efficiency depending on FMs}
To quantify the torque efficiency ($\xi$), the second harmonic Hall measurement was performed. In this measurement, a sinusoidal alternating current (AC) of 3 mA and 33.123 Hz was applied along the [1-10] direction of VN crystal, and the first (\(R_{xy}^{1\omega}\)) and second (\(R_{xy}^{2\omega}\)) harmonic Hall resistance was measured along [11-2] direction while rotating the in-plane magnetic field with respect to the current channel. An optical microscope image of the Hall bar device is shown in the inset of Fig.\ref{Fig3}(a). $R_{xy}^{1\omega}$ and $R_{xy}^{2\omega}$ can be expressed by the following equations \cite{yang2025large}
\begin{equation}\label{eq:1}
    R_{xy}^{1\omega}= R_{PHE}Sin2\varphi 
\end{equation}
where $ R_{PHE}$ is the planar Hall resistance. 
\begin{multline}\label{eq:2}
     R_{xy}^{2\omega} = Acos\varphi +  Bsin\varphi+ Ccos2\varphi +\\
    Dcos\varphi cos2\varphi + Esin\varphi cos2\varphi +FSin2\varphi+ G
\end{multline}
where $A$, $B$, and $C$ represent the $y$-, $x$-, and $z$-components of the damping-like (DL) torque, respectively. $D$, $E$, and $G$ correspond to the $y$-, $x$-, and $z$-components of the field-like (FL) torque, with $F$ additionally containing the contribution from the planar Nernst effect (see Supplementary Note 2 for full analysis expressions) \cite{yang2025large}. Figure\,\ref{Fig3}(a) shows out-of-plane magnetic field ($H_z$) dependent $R_{xy}^{1\omega}$ for VN(5)/[Co/Pt]$_3$, suggesting easy axis along [111] direction with amplitude ($R_{AHE}$) $\sim$ 1.7 $\Omega$. Figure\,\ref{Fig3}(b) shows the $R_{xy}^{1\omega}$ as a function of the in-plane magnetic field ($H_x$), from which the effective anisotropy field ($H_k^{eff}$) is found to be 10\,kOe. Figure\,\ref{Fig3}(c) presents the dependence of $R_{xy}^{1\omega}$ on azimuthal angle ($\varphi$) of $H_x$ with respect to the AC current direction. The fit using Eq.\,\ref{eq:1} revealed the $R_{PHE}$ of 0.12\,$\Omega$ as shown by the red line in Fig.\ref{Fig3}(c). Figure\,\ref{Fig3}(d) shows $\varphi$-dependence of $R_{xy}^{2\omega}$ (black), and the colored curve corresponds to the fits using individual torque components, while the red curve represents the full fit using Eq.\,\ref{eq:2}, which indicates the superposition of various torque components. We found that the DL torque components with $x$- and $y$-polarizations dominated the entire SOT in the VN(5)/[Co/Pt]$_3$, while the DL torque with $z$-polarization is much smaller. Note that the field-free magnetization switching is observed in the main sample, although the switching amplitude was partial as discussed in Fig.\ref{Fig5}(a).  Figure\,\ref{Fig3}(e) plots the value of -$2A/R_{AHE}$ as a function of $\frac{1}{H_x - H^{\mathrm{eff}}_k}$, to estimate the $y$-component of the DL field ($H^y_{DL}$) from the slope of linear fitting line. The slope of the fit yields the $y$-component of the DL effective field ($H_{DL}^y$), which is found to be -13.24\,Oe. The corresponding DL torque efficiency (\(\xi^y_{DL})\) is calculated using \cite{shukla2025berry}
\begin{equation}
\xi^y_{DL} = \frac{2e}{\hbar} \cdot \frac{ M_s t H_{DL}^y}{J_{\mathrm{VN}}}
\label{eq:5}
\end{equation}
where $M_s$, $t$, and $J_{VN}$ are the saturation magnetization, thickness of the FM layer, and the current density in the VN layer (see Supplementary Note 3 for estimation details). Using the experimentally determined $H_{DL}^y$, the resultant $\xi^y_{DL}$ was found to be -0.16 with $M_s$ $\sim$ 770 emu/cm$^3$ (see Fig.S7 (c) of supplementary Note), t $\sim$ 1.95 $\times$ 10$^{-7}$ and J$_{VN}$ = 4 MA/cm$^2$. Using same analysis we found $x$-component of the DL effective field  ($\xi^x_{DL})  = -0.059$. The $z$-component of the DL torque was found to be quite small, with a value of $\sim$ 0.005 at 15\,kOe. Similar procedure yielded the FL torque efficiencies of $\xi^y_{FL} = -0.14$ and $\xi^z_{FL} =- 0.03$. The $x$-component of FL torque efficiency ($\xi^x_{FL}$) is found to be small $\sim$ 0.012 at 15 kOe. Here, we discuss the $y$-component of DL torque, which plays a decisive role in magnetization switching from bulk SHE or OHE \cite{shukla2025berry,chiba2024comparative}. The $y$-component of effective spin–orbital Hall conductivity ($\sigma^{\mathrm{eff}}_{L-S}$= $\eta_{L-S}$$\cdot$$\sigma_{OHE}$) was calculated using $\tfrac{\hbar}{2e}\frac{\xi^{y}_{DL}}{\rho}$, where  $\rho$ $\sim$ 69\,\textmu$\Omega$\,cm is the resistivity of the VN layer (see supplementary Note 3). This yields a value of -1159 $\tfrac{\hbar}{e}$ S/cm. The corresponding spin-orbital Hall angle ($\theta_{L-S}$) was estimated to be $\theta_{L-S}$ $\sim$ $\frac{e}{\hbar}$ $\sigma^{eff}_{L-S}$/$\sigma_{xx}$ $\sim$ 0.08 with $\sigma_{xx}$ is 1.44 $\times$ 10$^{4}$ S/cm. This value is similar to $\theta_{SH}$ of Pt \cite{wang2014determination}. We also examined the $\xi^y_{DL}$ of VN/[Co/Pt]$_3$ with the VN thickness of 7.5\,nm and 10\,nm. For the VN thickness of 7.5\,nm, the slope of linear fit of -2A/$R_{AHE}$ versus $1/(H_x - H^{\mathrm{eff}}_k)$  (Fig.\,\ref{Fig3}(f)) gives the $H_{DL}^y$ of -27.96 Oe that corresponds to $\xi^y_{DL}$ to be -0.41 (see supplementary Note 4 for details). The $\sigma^{\mathrm{eff}}_{L-S}$ and $\theta_{L-S}$ is found to be -2135\,$\frac{\hbar}{e}$ S/cm and 0.15, respectively. For the 10\,nm VN device, we observed a reduced $\xi^y_{DL}$ of 0.04 (see supplemntary Note 4). The difference is possibly due to orbital current decay in thicker films or weak PMA (see supplementary Note 4). Notably, the $\xi^y_{DL}$ achieved in the 7.5\,nm device surpasses that of several heavy metals and light metal-based devices and is comparable to well-known W/CoFeB/MgO SOT devices \cite{kim2021spin,cha2021spin}.

To explore the FM layer that provides a large $\eta_{L-S}$ on the 5-nm-thick VN layer, the [Co/Pt]$_3$ in the main sample was replaced with Co(4 nm), Py(2 nm), and CoFeB(1.5 nm) layers, and second harmonic Hall measurements were conducted for each sample. The out-of-plane $R_{xy}^{1\omega}$ (Fig.\,\ref{Fig4}(a)) yields $R_\mathrm{AHE}$ = 0.58 $\Omega$ and $H_k^{\mathrm{eff}}$ = 17 kOe for the VN(5\,nm)/Co(4\,nm). Figure\,\ref{Fig4}(b) shows dependnece of $R_{xy}^{2\omega}$ on the $\varphi$ of $H_x$ (10\,kOe) with respect to current channel. The red line indicates the fitting curve using Eq.\ref{eq:2} to extract the coefficients.  Figure \ref{Fig4}(c) plots -$2A/R_{AHE}$ versus $\tfrac{1}{H_x+H_k^{\mathrm{eff}}}$ (black curve) and a linear fit yielded $H^y_{DL}$ = 1.65 Oe, corresponding to $\xi_{DL}^y$ = 0.08 and $\sigma^{eff}_{L-S}$ = 526 (\(\hbar/e\)) S/cm. Notably, this $\xi_{DL}^y$ is about half the value obtained for the VN(5\,nm)/[Co/Pt]$_3$ device. For FM = Py (Figs.\,\ref{Fig4}(d–f)) and CoFeB (Figs.\,\ref{Fig4}(g–i)) the $\xi_{DL}^y$ were estimated to be 0.015 and 0.002, respectively. 
\vspace{-2pt}
\begin{table}[htbp]
\centering
\caption{Summary of $R_{\mathrm{AHE}}$, $H_k^{\mathrm{eff}}$, $\xi_{\mathrm{DL}}$, and $\sigma^{\mathrm{eff}}_{\textit{L\text{-}S}}$ for the VN(5\,nm)/FM (=[Co/Pt]$_3$, Co, Py and CoFeB) systems.}
\begin{tabular}{lcccc}
\toprule
FM
& $R_{\mathrm{AHE}}$ (\(\Omega\))
& $H_k^{\mathrm{eff}}$ (kOe)
& $\xi_{\mathrm{DL}}$
& $\sigma^{\mathrm{eff}}_{\textit{L\text{-}S}}$ (\(\hbar/e\)) S/cm \\
\midrule
{[Co/Pt]$_3$} & $1.7$   & $10$     & $-0.16$ & $-1159$ \\
Co            & $0.58$  & $17$     & $0.08$  & $526$   \\
Py            & $0.24$  & $48.22$  & $0.015$ & $52$    \\
CoFeB         & $0.44$  & $41.5$   & $0.002$ & $8$     \\
\bottomrule
\end{tabular}
\label{tab:FM_properties}
\end{table}
The pronounced reduction compared with Co reflects the much lower $\eta_{L-S}$ of Py and the further suppression in amorphous CoFeB. The corresponding $\sigma^{eff}_{L-S}$ are determined to be 52 $\tfrac{\hbar}{e}$ S/cm and 8 $\tfrac{\hbar}{e}$ S/cm in case of Py and CoFeB FM layers. The value of $R_{\mathrm{AHE}}$, $H_k^{\mathrm{eff}}$, $\chi_{\mathrm{DL}}$, and $\sigma^{\mathrm{eff}}_{\mathrm{L\text{-}S}}$ for different samples is summarized in Table\ref{tab:FM_properties}. 

 \begin{figure*}[t]
   \centering
   \includegraphics[width=0.9\textwidth]{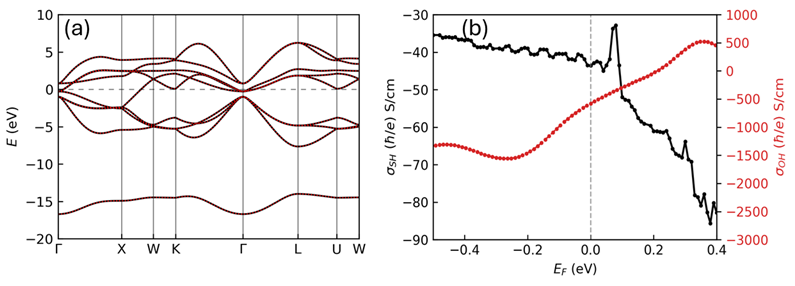}
   \caption{(a) The DFT (black curve) and Wannier (red) interpolated band structure of VN. (b) Fermi-level dependent spin Hall conductivity ($\sigma_{SH}$) and orbital Hall conductivity ($\sigma_{OH}$). The dashed line represents the position of the Fermi energy.}
    \label{Fig6}
\end{figure*}
\subsection*{Orbital torque driven magnetization switching}
To elucidate the effectiveness of orbital torque, current-induced magnetization switching (CIMS) is demonstrated for the VN(5)/[Co/Pt]$_3$, of which the magnetic properties are shown in Figs.\,\ref{Fig3}(a) and \ref{Fig3}(b). Figure \ref{Fig5}(a) shows the representative CIMS hysteresis loops with various $H_x$. The CIMS polarity reverses with the reversal of the sign of $H_x$. The $J_c$ for each CIMS as a function of $H_x$ is plotted in Fig.\ref{Fig5}(b) (red color). The $J_c$ excluding the current shunting into the [Co/Pt]$_3$ layer (see Supplementary Note 3) decreased with increasing $H_x$, which is attributed to the spin-orbital torque origin. The $J_c$ was estimated to be $\sim$ 18 MA/cm$^{2}$ at 300\,Oe. Note that field-free CIMS was observed with $J_c$ $\sim$ 22.5 MA/cm$^2$, although the values are larger than those with $H_x$, and the switching was not completed due to the nucleation of magnetic domains at the Hall cross patterns \cite{zhang2018size,wang2025large}. The CIMS for the VN(7.5)/[Co/Pt]$_3$ also measured (see supplementary Note 4), and the $J_c$ at each $H_x$ is shown in Fig.\ref{Fig5}(b) with blue color. The $J_c$ was found $\sim 14$ MA/cm$^2$ at 300\,Oe, which is 23\% smaller than the VN(5)/[Co/Pt]$_3$. The lower  $J_\mathrm{c}$ reflects the larger bulk-mediated orbital current in thicker VN, which is also evident from the increased $\xi_{DL}^y$ from second harmonic data as shown in Fig.\ref{Fig3}(f). We demonstrated CIMS in the W(5)/[Co/Pt]$_3$ for a comparison. The crystal structure and magnetic properties of the device are provided in the Supplementary Note 5. The CIMS switching is observed as shown in Fig.\ref{Fig5}(c), and the $J_c$ was lower but of the same order as the VN(5)[Co/Pt]$_3$ main sample. The variation of $J_c$ with $H_x$ is shown in Fig.\ref{Fig5}(d). This comparative study suggests the clear contribution of OHE in the VN layer to SOT, which is similar to the conventional SOT in the W layer, despite relatively weak SHE in VN. The power density, denoted as $P$ (= $J^2$$\rho$), of the VN(5)/[Co/Pt]$_3$ (VN(7.5)/[Co/Pt]$_3$) orbitronics device was determined to be 2.2 $\times$10$^{10}$ W/cm$^3$ (1.3 $\times$10$^{10}$ W/cm$^3$). This value is approximately half (one fourth) of that measured for the W(5)/[Co/Pt]$_3$ device, which exhibits a power density of 5.1 $\times$10$^{10}$ W/cm$^3$. These findings indicate a reduced incidence of Joule heating per unit volume in the VN-based devices compared to their W-based counterparts. Since the $J_c$ for the VN(7.5)/[Co/Pt]$_3$ is close to the W/[Co/Pt]$_3$, the reduced power consumption in the former device is attributed to its lower resistivity, which is an advantage in the light transition metal nitride systems. For the VN(10)/[Co/Pt]$_3$ device, the full switching could not be achieved even in the presence of an in-plane external field, which might be due to small $\xi_{DL}^y$ in the device (see supplementary Note 4).   

 Another characteristic in VN(5)[Co/Pt]$_3$ is the field-free CIMS. In general, the field-free CIMS arises from the unconventional spin current referred to as the collinear spin current, wherein the spin current flow is collinear with its polarization, such as a z-polarized spin current in a device exhibiting PMA \cite{han2024generation,zhang2024robust,yu2025field}. For instance, WTe$_2$ in its $T_d$ phase generates such a current through intrinsic mirror-symmetry breaking \cite{kajale2024field}. On the other hand, this characteristic in VN(5)/[Co/Pt]$_3$ was different from that reported so far, which is irrespective of the current direction, whether along the [1-10] mirror-symmetry direction or along the [11-2] broken mirror-symmetry direction (See Supplementary Note 6). The phenomenon suggests the absence of a $z$-polarized spin/orbital current resulting from the intrinsic symmetry breaking of VN, aligning with the analysis of second harmonic Hall measurement. Another possible interpretation could be the FM layer itself, i.e., tilting magnetization with respect to the out-of-plane direction. This situation is equivalent to the application of an in-plane magnetic field, which tilts the magnetization and disrupts the mirror symmetry perpendicular to the film plane. To check this for our VN(5)/[Co/Pt]$_3$, we examined the out-of-plane and in-plane magnetization and found an in-plane coercivity of 250\,Oe with tilting angle of 3.2$^\circ$ (see Supplementary Note 6 for detailed discussion). These factors fundamentally disrupt the mirror symmetry with respect to the film plane normal, thereby enabling field-free CIMS in the device. 

\subsection*{First principles calculation}
To quantitatively evaluate the SHC and OHC of VN, we performed first-principles calculations using density functional theory. The calculations were carried out within the experimentally observed lattice parameter and the cubic crystal symmetry of the Fm$\bar{3}$m space group. Figure \ref{Fig6}(a) displays the comparison between the Bloch bands (red) obtained from density functional theory and the Wannier-interpolated bands (black). The two sets of bands show excellent agreement across the entire energy window, confirming the accuracy of our Wannierization procedure—an essential prerequisite for reliable calculations of Berry-curvature driven transport properties. Notably, several band crossings appear close to the Fermi level, which can act as sources of spin and orbital Berry curvature and thereby give rise to the SHC and OHC in the system. The Fermi-level dependence of the $y$-component of SHC and OHC is presented in Fig.\,\ref{Fig6}(b). The magnitude of SHC ($\left| \sigma_{SH} \right|$, black curve) varies between 43 S/cm and 40 S/cm within an energy window extending 0.25 eV below the Fermi level. In contrast, the OHC ($\left| \sigma_{OH} \right|$, red curve) exhibits substantially larger values, ranging from about 600 S/cm to 1500 S/cm up to below 0.25eV. Therefore, the magnitude of the OHC in VN is several times larger than that of the SHC, indicating the dominant role of orbital transport in this light-element system.
 \begin{figure}[htbp]
   \centering
   \includegraphics[width=0.5\textwidth]{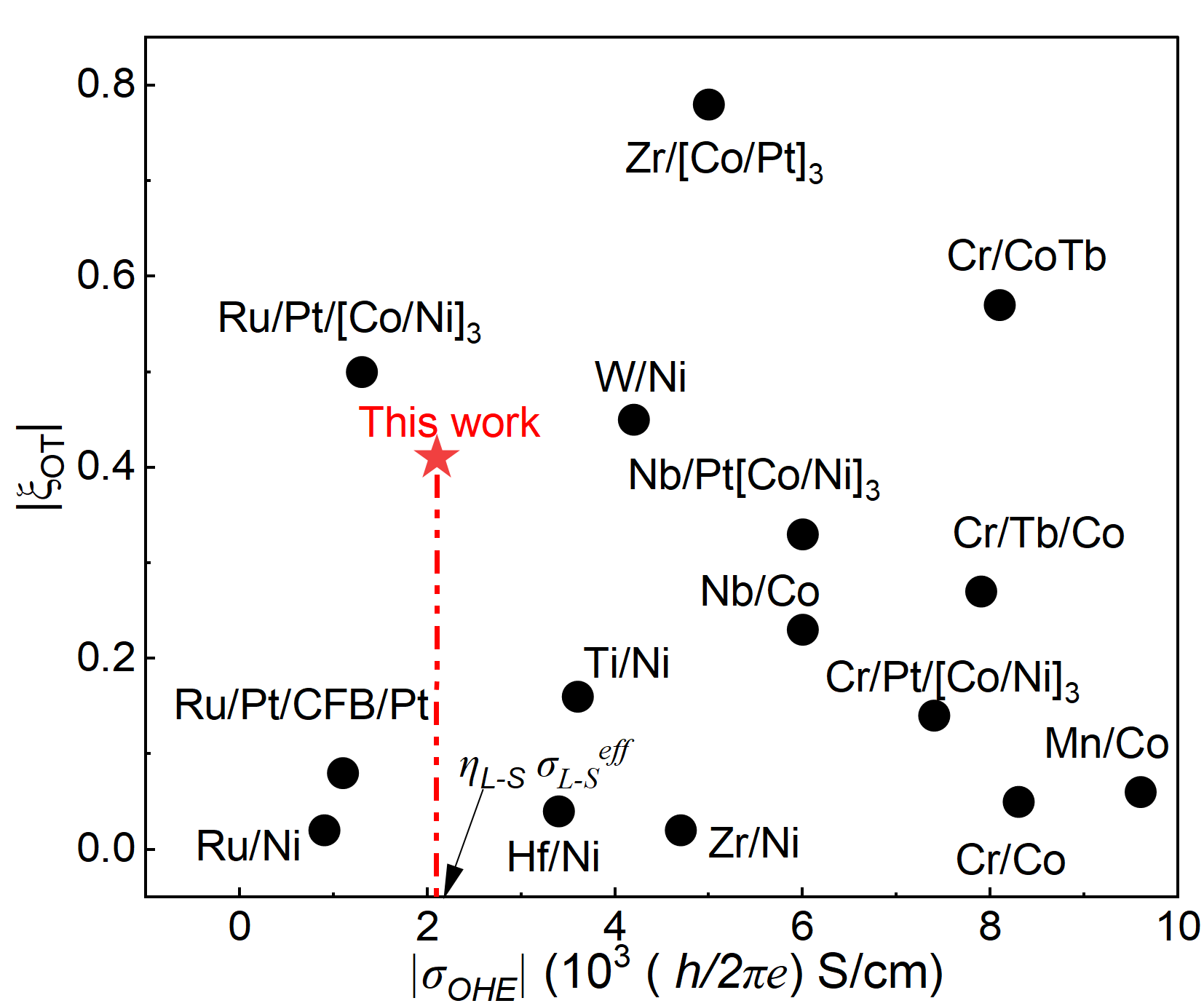}
   \caption{(a)  Comparison of the orbital torque efficiency versus orbital Hall conductivity for various material systems \cite{yang2024orbital}; the star indicates the values obtained in this work.}
    \label{Fig7}
\end{figure}
 
\vspace{-2mm}
\subsection*{Discussion}
The recently discovered phenomenon of orbitronics, which exploits the OHE of light metals, enables their use as orbital current sources in SOT devices. However, for the energy-efficient magnetization switching, the large OHE of LM and high $\eta_{L-S}$ of FM layer is essential. Furthermore, a perpendicular magnetized FM layer and its field-free switching are also desirable for the SOT device in terms of scalability and high-density memory. Most of the prior studies on orbital-based SOT devices have been conducted on the in-plane magnetized FM layer \cite{lee2021orbital,chiba2024comparative,liu2024harnessing}. Although such systems are indispensable for elucidating the underlying physical mechanisms, they are not viable for high-density memory devices. Among the several investigations on perpendicularly magnetized FM layers, most suffer from either low torque efficiency resulting in the high switching current density or high resistivity of the LM layer, which is undesirable \cite{lee2021efficient,yang2024orbital,gupta2025harnessing}. Here, we used a transition metal nitride, VN, with (111)-textured face-centred cubic structure exhibiting large OHE and low resistivity. This crystallographic texture further enables the epitaxial growth of a [Co/Pt]$_3$ multilayer FM stack that displays robust PMA and a large $\eta_{L\text{-}S}$ attributed to its strong SOC. As a result, we observed a large $\xi^y_{DL}$ of $-0.41$ in the VN(7.5)/[Co/Pt]$_3$ heterostructure, accompanied by a switching current density of 14 MA/cm$^2$. This performance is comparable to that of its spin counterpart W(5)/[Co/Pt]$_3$ in this study. The observed torque efficiency exceeds that reported for several recently studied SOT platforms (Fig.~\ref{Fig7} \cite{yang2024orbital}) and is comparable to the W/CoFeB-based SOT device \cite{kim2021spin,cha2021spin}. The $x$-axis value for our sample denotes the $\sigma^{eff}_{L-S}$ times $\eta_{L-S}$, since it is difficult to evaluate $\eta_{L-S}$ from the experiment.  Furthermore, we observed partial field-free CIMS in the VN/[Co/Pt]$_3$ irrespective of the current direction due to the finite in-plane magnetization of the [Co/Pt]$_3$ FM layer. Our work shows the potential of a hybrid bilayer structure of fully (111)-oriented transition metal nitrides and [Co/Pt]$_3$ for the orbital-based SOT devices. Nevertheless, the presence of Pt in our heterostructure raises concerns related to resource scarcity, increased magnetic damping, and elevated cost, which further suggest the discovery of the alternative FM layer with PMA and high $\eta_{L-S}$ for the sustainable SOT devices.
\section*{Conclusion}
In this work, a (111)-textured transition metal nitride VN layer has been used to switch the magnetization of a [Co/Pt]$_3$ FM layer by exploiting the OHE in VN and large $\eta_{L-S}$ of the FM. The VN(7.5\,nm)/[Co/Pt]$_3$ sample shows a large $\xi_{DL}$ of -0.41, which is comparable to or larger than several light metal-based SOT devices. Replacing [Co/Pt]$_3$ FM layer with FM=Co, Py, and CoFeB results in lower $\xi_{DL}$, suggesting orbital current from VN and its efficient conversion in [Co/Pt]$_3$ FM layer. The first principle calculation also suggests a large OHE in VN compared to its SHE. The CIMS measurement yields a $J_\mathrm{c}$ of 14 MA/cm$^2$ for the VN(7.5)/[Co/Pt]$_3$ heterostructure, which is comparable to that of its spin Hall-driven counterpart, W(5)/[Co/Pt]$_3$. Due to the relatively low electrical resistivity of VN, the power consumption in the VN(5\,,7.5)/[Co/Pt]$_3$ samples is reduced compared to that of the W(5)/[Co/Pt]$_3$. Our results highlight the promise of combining (111)-textured transition metal nitride with [Co/Pt]$_3$ FM layer having PMA for next-generation SOT devices by harnessing OHE from transition metal nitrides. 
\section*{Experimental details}
\textit{Film fabrication and characterization}: The Al$_2$O$_3$ (0001) substrate was cleaned in acetone and ethanol for 10 minutes, followed by flash annealing inside the sputtering chamber for 30 minutes under a base pressure of 1.2 $\times$ 10$^{-6}$ Pa at 600 $\degree$C. VN thin films were deposited by reactive DC magnetron sputtering using a gas mixture with N$_2$/(Ar+N$_2$) = 20\% at 450 $\degree$C. The Co/Pt, MgO, CoFeB, Py, and W layers were deposited by radio frequency (RF) magnetron sputtering at room temperature. The crystalline structure was characterized by XRD using Cu K$\alpha$ radiation (SmartLab; Rigaku Corporation), while the magnetic properties were measured with a magnetic property measurement system (MPMS; Quantum Design Inc.). 
\textit{CIMS and second harmonic Hall measurement}: A micro-sized Hall bar device with a length of 25\,\textmu m and width of 10\,\textmu m was fabricated using photolithography technique. The 5-nm-thick Ta and 50-nm-thick Au were deposited for the electrical contacts. A home-built system was used for CIMS measurement. A rectangular current of 0.5\,mA was applied to the current channel of the Hall bar device with a duration of 10\,ms using a pulse generator. A sense current of 0.5\,mA was applied to measure the Hall resistance 1\,s after each current pulse. The harmonic Hall resistance was measured using a lock-in amplifier (LI5640; NF Co.), while $H_x$ and $\varphi$ were scanned.  A sinusoidal current of 3.0\,mA (4\,MA/cm$^2$) with a frequency of 33.123\,Hz was applied using a pulse generator (FG420; Yokogawa Electric Co.). All measurements were performed at room temperature. 

\textit{Procedure for calculation of spin and orbital Hall conductivity}: The electronic structure and transport properties were calculated using density functional theory using Quantum ESPRESSO \cite{giannozzi2009quantum}, and the Vienna \textit{ab initio} Simulation Package (VASP) \cite{hafner2008ab} was employed to compute SHC and OHC, respectively. The Perdew–Burke–Ernzerhof (PBE) form of the generalized gradient approximation (GGA) was used for the exchange-correlation functional \cite{ernzerhof1999assessment}. A plane-wave basis set with an energy cutoff of 80\,Ry was applied. The Brillouin zone was sampled using an 8\,$\times$8\,$\times$8 $k$-point mesh. The SHC and OHC were evaluated using linear response theory \cite{nakano1957method,kubo1957statistical}.
\begin{equation}
\sigma^{X(l)}_{\alpha\beta}(E) = \frac{e}{V} \sum_{\mathbf{k}} 
\Omega^{X(l)}_{\alpha\beta}(\mathbf{k},E)
\label{eq3}
\end{equation}

$\Omega^{X(l)}_{\alpha\beta}(\mathbf{k},E)$ is the orbital Berry curvature provided by \cite{miura2021first}.

\begin{align}
\Omega^{X(l)}_{\alpha\beta}(\mathbf{k},E)
= {} & 2\,\frac{\hbar^{2}}{m_{e}^{2}}
\sum_{n \neq m} 
\frac{
\mathrm{Im}\!\left[
\langle k n | \hat{\rho}^{X(l)}_{\alpha} | k m \rangle
\langle k m | \hat{\rho}_{\beta} | k n \rangle
\right]
}{\left( \varepsilon_{kn} - \varepsilon_{km} \right)^{2}}
\nonumber \\
& \times \left[ f_{kn}(E) - f_{km}(E) \right].
\label{eq4}
\end{align}
where $V$ denotes the unit-cell volume, $m_e$ is the mass of the electron. The $m$ and $n$ are the occupied and unoccupied band indices, respectively. The operator $\hat{\rho}^{X(l)}_{\beta}$ denotes the “spin’’ or “orbital’’ current operator ($X =$ spin or orbital), where  
\[\hat{\rho}^{\mathrm{spin}(l)}_{\alpha} = \hat{p}_{\alpha}\,\hat{S}_{\gamma} 
+ \hat{S}_{\gamma}\,\hat{p}_{\alpha},\qquad\hat{\rho}^{\mathrm{orbital}(l)}_{\alpha} 
= \hat{p}_{\alpha}\,\hat{L}_{\gamma} + \hat{L}_{\gamma}\,\hat{p}_{\alpha}.
\]
Here $\hat{p}_{\alpha}$ ($\hat{p}_{\beta}$) is the momentum operator along the $\alpha$ ($\beta$) direction.,  $\hat{S}_{\gamma}$ and $\hat{L}_{\gamma}$ denote the spin and orbital angular momentum operators along the $\gamma$ direction, respectively.  $\lvert k n \rangle$ is the eigenstate with energy $\varepsilon_{kn}$, and $f_{kn}(E)$ gives the occupation probability of band $n$ and wave vector $\mathbf{k}$  at energy $E$ relative to the Fermi energy. The $\sigma_{OH}$ of VN was computed using a  $15 \times 15 \times 15$ $k$-point mesh in the first Brillouin zone.



\section*{Acknowledgments}
This work was supported by KAKENHI Grants-in-Aid
No. 23K22803 from the Japan Society for the Promotion of
Science (JSPS). Part of this work was carried out under the
Cooperative Research Project Program of the RIEC, Tohoku
University.


The authors declare no conflicts of interest.
\clearpage
\bibliographystyle{MSP}
\bibliography{References}

\end{document}